\documentclass[9pt,twocolumn,twoside]{opticajnl}
\journal{opticajournal} 

\setboolean{shortarticle}{true}

\usepackage{lineno}

\title{SOA-based reservoir computing using up-sampling}

\author[1,*, $\dag$]{E. Manuylovich}
\author[2, *]{A.E. Bednyakova}
\author[2]{D.A. Ivoilov}
\author[3]{I.S. Terekhov}
\author[1]{S.K. Turitsyn}

\affil[1]{Aston Institute of Photonic Technologies, Aston University, Birmingham B4 7ET, UK}
\affil[2]{Novosibirsk State University, Novosibirsk 630090, Russia}
\affil[3]{School of Physics and Engineering, ITMO University, St. Petersburg 197101, Russia}
\affil[*]{These two authors contributed equally to this work}
\affil[$\dag$]{e.manuylovich@aston.ac.uk}

\begin{abstract}
We introduce a new approach to reservoir computing based on up-sampling and modulation, utilizing semiconductor optical amplifier and photodetector as nonlinear elements without conventionally used delay loop. We demonstrated the 400-step prediction capability of the proposed scheme for the Mackey-Glass time series test.
\end{abstract}

\setboolean{displaycopyright}{false} 

\begin{document}

\maketitle

\section{Introduction}
Reservoir Computing (RC) is a general approach in the domain of neural network architectures that has two key components: (i) a non-trainable recurrent neural network with random weights as a "reservoir", and (ii)  training is done using only linear readout. The advantage of RC is that the network training comes down to a simple linear regression, which is typically fast and straightforward. The reservoir computing implementation in ultra-fast nonlinear photonic systems has attracted a great deal of new attention recently (see, e.g., \cite{brunner2018tutorial,van2017advances,tanaka2019recent,appeltant2011information,sorokina2019fiber,sorokina2020multidimensional} and references therein).

The first wave of implementations of RCs has been focused on mimicking the mathematical model by designing special optical or optoelectronic systems governed by master equations similar to the targeted models. 
An interesting new direction in the generalization of original RC approach and practical implementation of such modified concepts is to use given/chosen physical systems for computing \cite{zhang2014integrated,sunada2019photonic,marcucci2020theory,wright2022deep} that can also be considered as an extreme learning approach \cite{huang2006extreme} if no recurrent connections are employed \cite{ortin2015unified}. In particular, nonlinear optical wave systems can offer high entropy of the phase space that can be utilized as a computational resource \cite{sunada2019photonic,marcucci2020theory}. 

 One of the popular implementations of RC is a delay-reservoir computer, which includes a single nonlinear node and delay line \cite{brunner2018tutorial}. Application of the semiconductor optical amplifier (SOA) as a nonlinear node in systems with a delay has already been actively studied for some years (see, e.g., \cite{appeltant2011information,van2017advances,vandoorne2011photonic,brunner2018tutorial} and references therein). Here, we propose a new approach using a single nonlinear device – SOA without delay lines to perform computations. Combining up-sampling (the time slot for each symbol is sampled with a high rate compared to the baud rate) with signal propagation through SOA, we transform the input nonlinearly into a high-dimensional space. After the photodetector (that is effectively an additional nonlinear transformation), the features of the output vector can be read out digitally and used in the learning algorithm. In contrast to \cite{takano2018compact}, which used a faster output sampling rate but could not leverage its benefits due to the low bandwidth of the readout system, our approach demonstrates the advantage of upsampling in enhancing the feature matrix rank. Additionally, unlike the delay-based reservoir computers in \cite{takano2018compact, nakajima2021scalable}, our method employs a delay-less approach while providing comparable memory capacity.

\section{SOA-based reservoir computing scheme}

\begin{figure*}[ht]
\centering\includegraphics[width=\linewidth]{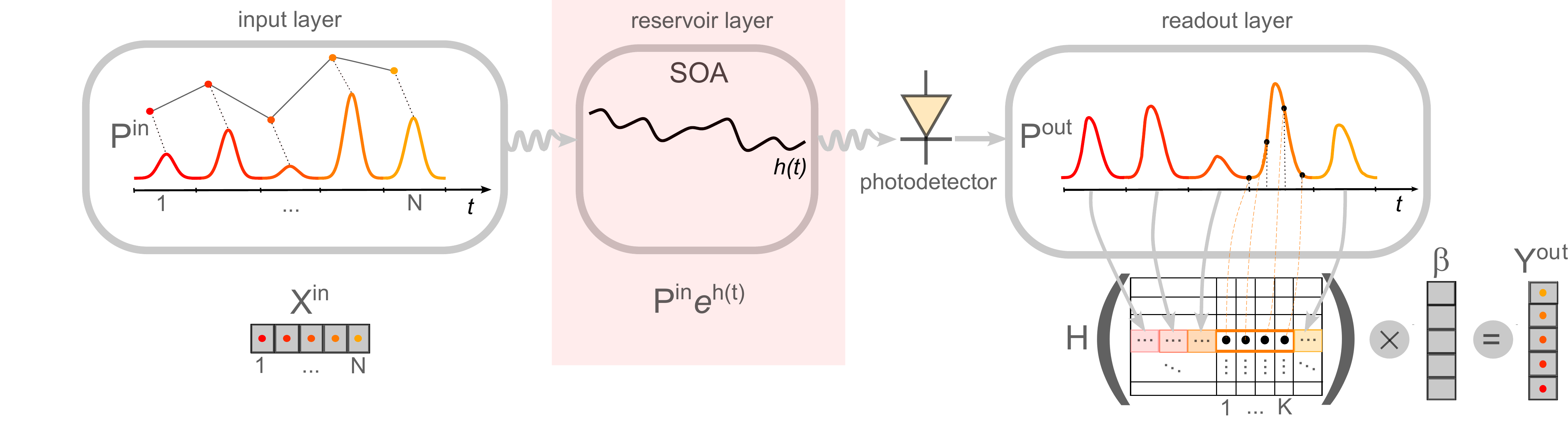}
\caption{The basic SOA-based network architecture for reservoir computing.}
\label{fig: RC}
\end{figure*}

The master model for SOA describes how the input optical field $A_{in}(t)= \sqrt{P_{in}(t)}\, \exp( i \phi_{in}(t))$ is transformed into the output field $A_{out}(t)= \sqrt{P_{out}(t)}\, \exp( i \phi_{out}(t))$ \cite{agrawal1989self}:
\begin{gather}
P_{out}(t)= P_{in}(t) \exp[h(t)], \\
\phi_{out}(t)=\phi_{in}(t)-\frac{\beta}{2} h(t),\\
\frac{d h}{d t}=-\frac{h-h_0}{\tau_c}-\frac{P_{in}(t)}{E_{sat}}\left[\exp(h)-1\right],\label{soa}
\end{gather}
where in/out index means the input/output signal, $\beta$ is the linewidth enhancement factor or Henry factor,  $h_0$ is the integral small-signal gain, $\tau_c$ is the gain recovery time, $E_{sat}$ is a characteristic saturation energy. Without loss of generality, we will use in numerical modeling: $\beta=5$, $h_0 = 30$ dB, $\tau_c = 200$ ps, $E_{sat}=8$ pJ.
We exploit the nonlinear transformation of the input signal stream by SOA with the memory introduced by the delayed gain recovery to perform computations. It is seen that, in general, SOA manifests memory: gain at a certain point in time
$h(t)$ depends on the signal in the previous moments.

The proposed scheme, shown in Fig.~\ref{fig: RC}, includes an input layer formed by a train of optical pulses with the modulated power or amplitude, SOA as a nonlinear reservoir, followed by a photodetector converting optical field $A_{out}(t)$ into an electrical signal $I(t)=\kappa P_{out}(t)=\kappa |A_{out}(t)|^2$, which is used for the output readout layer. Applying linear re-scaling, we can set $\kappa =1$ and use $P_{out}(t)$  as the output function. Consider, without loss of generality, the following simple amplitude encoding of the input vector $X^{in} = (x^{in}_1, x^{in}_2, ..., x^{in}_N)$: 

\begin{equation}
 A_{in}(t)= \sqrt{P_0}\sum_{n=1}^{N} c_n\, g\left(t-\left(n-\frac{1}{2}\right) T_s\right),
\label{eq: symbol encoding via pulse train}
\end{equation}
where $g(t)$ is a carrier pulse shape, e.g., we consider here the Gaussian optical pulses: $g(t) = \exp[-t^2/(2T_0^2)]$, $N$ - number of the pulses, $T_s$ - time slot or symbol interval, $P_0$ - normalization power parameter, $c_n$ is amplitude
modulation parameter for $n$-th slot that is linked to the input vector $X^{in}$: $x^{in}_n = c^{2}_n$. 

We measure $K$ samples per each transmitted pulse, resulting in $N\times K$ points for each pulse train. The sampling rate $f$ was set to 80 GSa/s to match the typical performance of a good oscilloscope. The symbol interval $T_s$ was equal to 200 picoseconds, and the number of samples per pulse $K$ was equal to 16, matching the sampling rate. 

Using the nonlinearly transformed signal at the output of the SOA, we can build a vector of features used for forecasting:

\begin{equation}
P^{out}_k=P_{in}(t_k)\exp\left[h(t_k)\right]\bigg\rvert_{t_k= k/f}, k=1,..., N\times K
\end{equation}

After passing the pulse train through the system, we can write the obtained feature vector as a corresponding row in the feature matrix $\mathbf{H}$. The number of virtual neurons in the output layer is $KN$, where $N$ is the number of symbols used for prediction and $K$ is the number of samples per symbol due to upsampling. In the exploration of the memory capacity of the SOA-based RC, we put only the last $K\cdot(N-q)$ samples of the feature vector explicitly to the feature matrix, and the first $K\cdot q$ of it only implicitly affects the entities of $\mathbf{H}$ via gain saturation. In what follows, we consider a general case with $q$ symbols in the pre-sequence.
When we teach the SOA-based RC to forecast a sequence, we pass $M$ multiple pieces of pulse sequences of length $N$ through the system and measure the nonlinearly mixed output feature vector of length $K\cdot(N-q)$. So we construct feature matrix $\mathbf{H}$ using these $K\cdot(N-q)$ readout channels linearly distributed over each of the $N-q$ time slots: $\mathbf{H} \in \mathbb{R}^{M \times K\cdot(N-q)}$ ($M$ rows and $K\cdot(N-q)$ columns). We define this feature matrix as:
\begin{equation}
H_{m,k}=P^{out}_{m,k},  k=  K\cdot q+1,..., N\times K
\end{equation} 

When we use the proposed SOA-based RC for the forecasting problem, we do it in the form of linear regression, i.e., the output vector $Y^{out}\in \mathbb{R}^{M\times 1}$ is assigned to the feature matrix $\mathbf{H}$ as a linear combination of its columns:
\begin{equation}
\mathbf{H} \vec{\beta} = Y^{out}
\label{eq:regr}
\end{equation}
where vector $\beta\in \mathbb{R}^{K\cdot(N-q)\times1}$ - the vector of the output weights, which can be found from eq.~(\ref{eq:regr}) using ridge regression or taking pseudo-inverse of matrix $\mathbf{H}$.

To measure the accuracy of the forecast, we consider mean absolute scaled error (MASE) in all the tasks \cite{HYNDMAN2006}: 
\begin{equation}
MASE=\frac{\frac{1}{M}\sum_{m=1}^M|Y^{t}_{m}-Y^{out}_{m}|}{\frac{1}{M-1}\sum_{m=2}^M|Y^{t}_{m}-Y^{t}_{m-1}|},
\label{eq: mase}
\end{equation}
where $Y^{t} \in \mathbb{R}^{M\times 1}$ is the actual data. Eq.~\ref{eq: mase} can be understood as the ratio of the mean absolute error of the forecasted values to the mean absolute error of the in-sample one-step naive forecast. $MASE>1$ suggests that the actual forecast should be discarded in favor of a naive forecast. We chose this metric for its interpretability, as it clearly indicates whether the proposed forecasting algorithm outperforms a simple naive prediction. This metric is well-suited for model comparison and determining optimal parameters.

\subsection{Linear model with nonlinear photodetector}

In the linear model in the absence of SOA, the elements of the matrix $H$ can be represented as:

\begin{equation}
    H_{m,k}=P_0 \sum_{l=q}^n\sum_{p=0}^n\sqrt{x_{m,l} x_{m,p}}V_{k,l,p},
    \label{eq: linear analytic}
\end{equation}
where 
\begin{equation}
V_{k,l,p}=g\left(k/f-(2l+1)T_s/2\right)g\left(k/f-(2p+1)T_s/2\right)
\label{eq: gaussian pulses product}
\end{equation}
These equations can be obtained by taking an absolute value squared of the encoded complex amplitude given by Eq. \ref{eq: symbol encoding via pulse train} and collecting pairwise products of Gaussian functions.
The solution $\vec{\beta}$ can be written in the following form:
\begin{eqnarray}
    \vec{\beta}=H^{\dag} \vec{Y}^{t},
\end{eqnarray}
where $H^{\dag}$ is the Moore–Penrose inverse matrix. Due to the nonlinearity of the detector, with an appropriate choice of parameters, symbols can be mixed, and this simplest system (without SOA) can potentially also be used for computing. Therefore, we compare cases with and without SOA to clarify the photodetector's effect as a nonlinear transformation for the considered signal and system parameters.

\section{Results}

\begin{figure}[t!]
\centering
\includegraphics[width=0.35\textwidth]{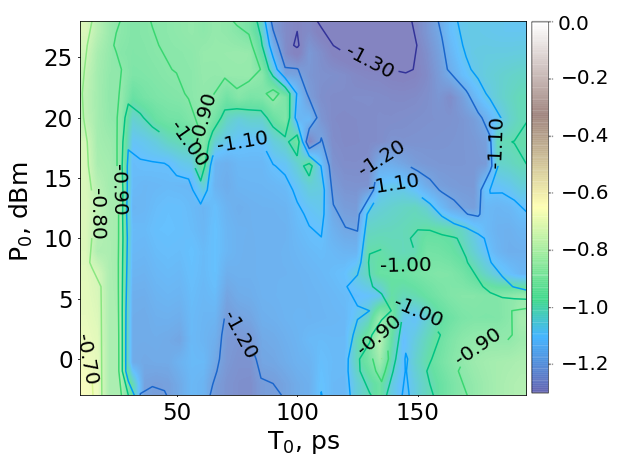}
\caption{MASE (in logarithmic scale) obtained for one-step prediction in the plane ($T_0$, $P_0$). A single previous symbol is used to predict the next one, pre-sequence length $q=2$.}
\label{fig: MG_map}
\end{figure}

The performance of the proposed scheme was analyzed using the benchmark prediction task of learning the Mackey-Glass (MG) time series for chaotic attractor:
\begin{equation}
\frac{dy(t)}{dt}=\frac{\alpha y(t-\tau)}{1+y(t-\tau)^{\beta}}-\gamma y(t),   
\end{equation}
where $\alpha=0.2$, $\beta=10$, $\gamma=0.1$, $\tau = 17$. The time series was generated by numerically integrating the MG equation with the fourth-order Runge-Kutta method and a time step $\Delta t = 0.1$. 
Then, the numerical solution calculated at discrete equally spaced times was scaled by a transformation $y \mapsto (y - y_{min})/y_{max} + 0.1$. 
The 0.1 shift prevents negative values during prediction. 
The resulting data was downsampled by a factor of 10, resulting in a time series with 3600+2$N$ data points.
Subsequently, we employed a sliding window technique on the time series, creating 3600 shorter data sequences, each with a length of $N$ so that the first sequence contains symbols from $x_1$ to $x_N$, the second one contains symbols from $x_2$ to $x_{N+1}$, and so on.
These sequences were evenly divided into two sets: the training set $\mathbf{X}^{in}_{train} \in \mathbb{R}^{M \times N}$ and the testing set $\mathbf{X}^{in}_{test} \in \mathbb{R}^{M \times N}$ ($M=1800$).
Each of the $M$ data sequences was then encoded by the peak amplitude of the Gaussian pulse, as described in the previous section. 

In the first step, we investigate the memory capabilities of the SOA, using a single MG symbol to compute the weights for predicting one step ahead. More precisely, after a pulse train of length $N$ propagates through the SOA, we record only the last $K$ samples in the row of the feature matrix. 
Figure~\ref{fig: MG_map} shows the prediction error on a two-dimensional plane, representing two optimization parameters: pulse duration $T_0$ and normalization power parameter $P_0$. When the pulse sequences include a single symbol that goes into the feature matrix and two symbols in the pre-sequence, the prediction error is close to zero. Prediction forward from features generated from a single symbol without any memory is impossible, as the system cannot determine whether the next symbol should increase or decrease based solely on the current symbol. Therefore, the fact that the prediction error is close to zero indicates that the system possesses some memory. This ability to "remember" past symbols, which are not accounted for in the feature matrix, can be attributed to SOA memory, where the amplification of the current pulse depends on prior pulses.

\begin{figure}[h!]
\centering
\includegraphics[width=0.4\textwidth]{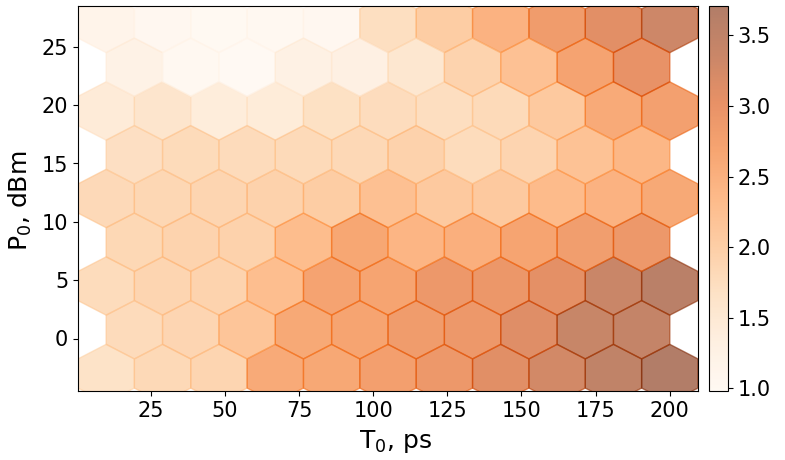}
\caption{Memory capacity MC depending on pulse duration $T_0$ and peak power $P_0$. Pulse spacing $T_s=200 ps$, and $K=16$.}
\label{fig: MC}
\end{figure}

\begin{figure}[!ht]
\centering
\includegraphics[width=0.35\textwidth]{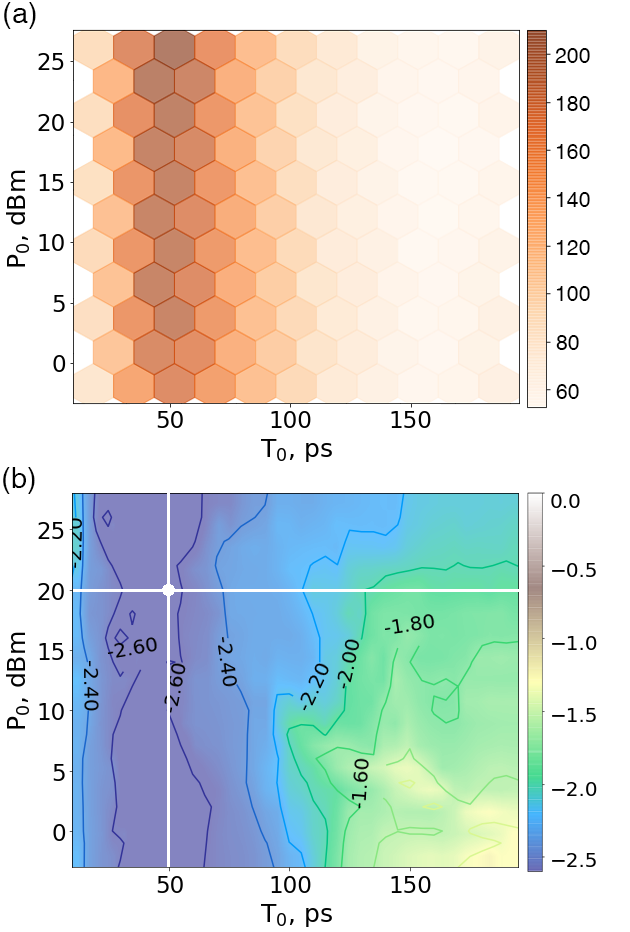}
\caption{Feature matrix rank (a) and MASE in logarithmic scale (b) in the plane of the pulse parameters ($T_0$, $P_0$). Fifteen previous symbols are used to predict the next one. $N=15$, $q=0$.}
\label{fig: MG_15}
\end{figure}

We further investigated the memory capacity of the proposed approach. The memory capacity $MC$ is defined as \cite{ortin2015unified}:
\begin{equation}
MC = \sum_{i=1}^{\infty} \frac{\left< (y(n-i))(o_i(n)) \right>^2}{\sigma^2(y(n-i)) \sigma^2(o_i(n))}
\end{equation}
Here, $y(n)$ is a random input signal, $o_i(n)$ is the reservoir output at time $n$ with output weights trained on the $i$-th past input signal $y(n-i)$, $\sigma^2$ is the variance, and $\left< \right>$ denotes the time average \cite{ortin2015unified}.
The memory capacity $MC$ quantifies how much past input signal information can be reproduced. We used a feature matrix composed of a single symbol with upsampling $K=16$ and tested prediction on MG series with pre-sequence length $q=10$. The result is shown in Fig. \ref{fig: MC}.
One can see that higher power levels lead to shorter memory. We believe it is because even a single high-power pulse can saturate the SOA gain and "erase" any traces from previous pulses by setting $h$ to near zero. For lower power levels, $MC>3.5$, which exceeds the $MC$ of some even more complicated approaches ($MC<1.6$ in \cite{takano2018compact}), but is outperformed by some of the larger-scale delay-based RC schemes with $MC=8$ \cite{ortin2015unified}.
The symbol memory capacity can be improved by either using SOA with a longer carrier lifetime or by increasing the symbol rate. However, pulse parameters which provide longer memory capacity have lower peak power and gain nonlinearity, reducing the effective dimensionality of the RC's feature space. Figure \ref{fig: MG_15} shows the feature matrix rank based on these pulse parameters.

On the other hand, an increase in the length of the pulse sequence, along with the number of symbols used for predicting the next one, results in a significant improvement in forecast accuracy (see Fig.S1 in supplementary materials). 
Fig.~\ref{fig: MG_15}(b) depicts MASE in the plane $(T_0, P_0)$, calculated for a pulse train length of $N=15$ where all of the pulses are used for prediction. The figure shows that the prediction error exhibits a localized minimum at approximately $T_0=50$ ps. This minimum corresponds to the points in Fig.~\ref{fig: MG_15} at which the feature matrix $H$ achieves its maximum rank.

\begin{figure}[h!]
\centering
\includegraphics[width=0.35\textwidth]{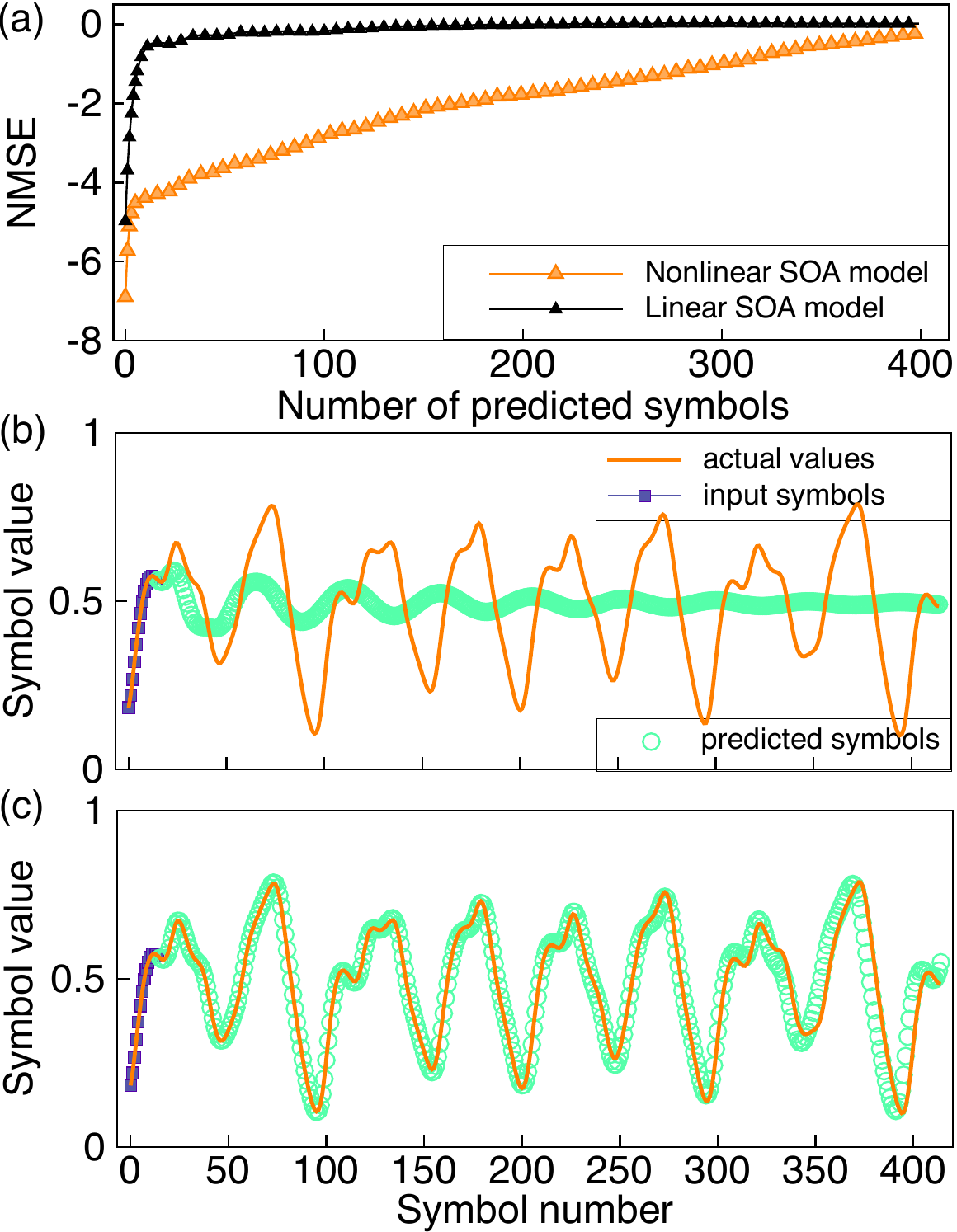}
\caption{(a): Comparison of testing prediction errors between the linear model with a nonlinear detector and nonlinear SOA model for multi-symbol prediction. (b)-(c): Results of 400-steps ahead MG prediction for the linear model with nonlinear detector (b) and nonlinear SOA model (c). $T_0=50$ ps, $P_0=20$ dBm, $N=15$, $q=0$.}
\label{fig: MG}
\end{figure}

Next, we examine multi-step Mackey–Glass forecasting for encoding parameters $T_0=50$ ps and $P_0=20$ dBm indicated by the white circle in Figure~\ref{fig: MG_15}b. To predict future values of the time series, we employ a recursive strategy (see Fig.~\ref{fig: RC}). This strategy involves feeding the results of predictions at each time step as input to the network for the next time step. The forecasting results for 400-steps-ahead Mackey-Glass prediction are presented in Fig.~\ref{fig: MG}. To compare the results with existing RCs, we employed the normalized mean squared error (NMSE) to measure the accuracy. Compared to other papers on photonic RC for Mackey-Glass prediction, our approach outperforms \cite{dong2019optical} with an NMSE of \textasciitilde0.01 for predicting 300 symbols (6 quasi-periods) versus \textasciitilde0.1 in \cite{dong2019optical}. When compared to delay-based RC \cite{ortin2015unified}, our method shows similar accuracy for single-step prediction, but multi-step prediction accuracy cannot be compared due to the absence of error metrics in \cite{ortin2015unified}. We compare two models: the linear model with nonlinear detector (Fig.~\ref{fig: MG}b) and the nonlinear SOA model (Fig.~\ref{fig: MG}c). Note that Figs.~\ref{fig: MG}b-c display forecasts that correspond to one of $M=1800$ data sequences in the testing set, each having a length of $N=15$. These 15 symbols, used for the prediction of the following 400 symbols, are shown in purple in the figure. While the linear model is capable of making forecasts for several steps ahead, it is not suitable for longer prediction horizons and can only capture basic periodic MG dynamics (Fig.~\ref{fig: MG}b). By incorporating an amplifier into the system, we can expand the feature space dimension and considerably extend the prediction horizon.

In conclusion, we proposed a new approach to reservoir computing based on optical signal up-sampling and modulation, exploiting semiconductor optical amplifier and photodetector as nonlinear elements. We showed that a delay line is not necessary for RC, and the concept of RC can be implemented without any delay lines. We demonstrate that in the same way that the attention mechanism is sufficient for translation \cite{vaswani2017attention}, nonlinearity and internal SOA dynamics are sufficient for RC. Using traditional
Mackey-Glass time series was used as a test, and we demonstrated excellent prediction capability of the proposed scheme for 400 steps.

\section{Acknowledgments}
The work of DAI and AEB was supported by the Russian Science Foundation (Grant No.21-42-04401).
EM and SKT acknowledge the support of the EU H2020 ITN project POSTDIGITAL (No. 860360) and the Engineering and Physical Sciences Research Council (project EP/W002868/1).

\begin{backmatter}

\bmsection{Disclosures} The authors declare no conflicts of interest.

\bmsection{Data availability} Data underlying the results presented in this paper may be obtained from the authors upon reasonable request.

\bmsection{Supplemental document} See Supplement 1 for supporting content.

\end{backmatter}

\bibliography{references}

\bibliographyfullrefs{references}

\end{document}